# Magnetic-field cycling induced anomalous irreversibility in resistivity

# of charge-ordered manganites

K. R. Mavani and P. L. Paulose

*Dept. of Condensed Matter Physics and Materials Science, Tata Institute of Fundamental Research, Mumbai – 400 005, India*

## ABSTRACT

The rare-earth ions (RE = Eu, Dy Ho, Tm, Y) substituted charge-ordered antiferromagnetic manganites, $Pr_{0.45}RE_{0.05}Ca_{0.5}MnO_3$, were studied for the magnetic and the transport properties in the presence of external magnetic-fields of up to 14 Tesla. Regardless of the intrinsic magnetic property of RE ions, all the compounds exhibit successive step-like metamagnetic transitions at low temperatures, which are strongly correlated to their electronic transitions. At any fixed temperature in two different temperature-regimes, we observed contrary effects of the magnetic-field cycling on the resistivity of these manganites, namely, i) in the low temperature regime (≤70 K), the resistivity was *irreversible* showing *lower* values than initial after a magnetic-field cycle was over, which is consistent with the *irreversible* magnetization, and ii) in a temperature regime above 70 K, the resistivity is *irreversible* with noticeably *higher* values than initial, whereas the magnetization was found to be *reversible*. For the latter case, we further show that this irreversibility of resistivity systematically depends on the temperature and the magnitude of applied magnetic-field. These results suggest that the observed resistivity behavior originated from the magnetic-field induced metamagnetic transitions and training effect.

The $ABO_3$ type perovskite manganites have attracted considerable attention of researchers in the recent past and are extensively studied for their fascinating properties such as negative colossal magnetoresistance, orbital-charge ordering, and correlated electronic and magnetic transitions.[1] The study on these manganites was extended by the observations of metamagnetic transition (MMT) with a concomitant melting of charge-ordered state and the phase-separation[2-4]. One of these, the half-doped $Pr_{0.5}Ca_{0.5}MnO_3$ (PCMO) shows a robust charge-exchange type charge-ordered (CO) antiferromagnetic (AFM) state that melts to a ferromagnetic (FM)-metallic state on application of a high critical magnetic-field ($H_c$) of about 25 T at low temperatures.[2] An '$A$-site' cation disorder[5] $\left(\sigma^2 = \sum x_i r_i^2 - \langle r_A \rangle^2\right)$ induced by fractional doping of $Ba^{2+}$ weakens its CO state.[6] This is evident from the reduction of the $H_c$ by almost an order of magnitude in fractionally $Ba^{2+}$ substituted PCMO. Also, a systematic shifting of $H_c$ has been observed with increasing $\sigma^2$.

Thus far, reduced $H_c$, weakened CO state, and metamagnetism in $Pr_{0.5}Ca_{0.5}MnO_3$ system have been achieved either by Mn-site doping or by 'larger' isovalent cation substitution at $A$-site.[4,6-8] However, there is no report on the isovalent 'smaller' ion doping at $A$-site in PCMO system. In the present study, we have successfully obtained the lower $H_c$ in smaller rare-earth ions (RE = $Eu^{3+}$, $Dy^{3+}$, $Ho^{3+}$, $Tm^{3+}$ and $Y^{3+}$) doped $Pr_{0.45}RE_{0.05}Ca_{0.5}MnO_3$ manganite system as a result of cation disorder effect, importantly, with almost constant tolerance factor $\left(t = (\langle r_A \rangle + r_O)/\sqrt{2}(r_{Mn} + r_O)\right)$ and a fixed $Mn^{3+/4+}$ ion ratio (1:1) in this system. The cation disorder may influence a phase-separated state,[9,10] which is particularly interesting as there are numerous interfaces between ferromagnetic (FM) and antiferromagnetic (AFM) phases; and the exchange interactions across the phase boundaries affect the magnetic and the electrical properties. For instance, due to repeated thermal cycling or magnetic-field cycling across the magnetic transition, so-called 'training effect' has been observed in ferromagnetic or metamagnetic manganites, where both the magnetization and the resistivity were consistently irreversible.[11-15] In this context, the presently studied $Pr_{0.45}RE_{0.05}Ca_{0.5}MnO_3$ manganite system exhibits atypical resistivity behavior, namely, i) at low temperatures (below 70 K), the magnetic-field dependent magnetization and resistivity isotherms are consistently irreversible and ii) in the high temperature regime (80-200 K), the field-dependent magnetization is reversible in contrast to irreversible resistivity. We have explored the latter temperature regime, and in this letter, we report the magnetic-field cycling induced anomalous irreversibility in the resistivity isotherms in applied fields of up to 14 T. The results are discussed in the framework of magnetic-field cycling induced training effect.

The $Pr_{0.45}RE_{0.05}Ca_{0.5}MnO_3$ (RE = Y, La, Eu, Dy, Ho, Tm) compounds were synthesised by conventional solid state reaction method. The high purity ($\geq$99.95%) powders of $Pr_6O_{11}$, $RE_2O_3$, $CaCO_3$ and $MnO_2$ were taken in the stoichiometric proportions. These powders were mixed and ground thoroughly. The mixtures of powders were heated several times at temperatures between 950°C to 1250°C with intermediate grindings and finally sintered in pellet form at 1350°C for 24 h. The X-ray diffraction (XRD) data were recorded using X'Pert powder diffractometer and analysed by Rietveld method (using 'FULLPROF' programme code). The magnetization measurements were carried out using vibrating sample magnetometer (Oxford) and the magnetoresistance was measured with four-probe method in varying applied magnetic-fields up to 14 T using physical property masurement system (Quantum Design). In the further discussions, the $Pr_{0.45}RE_{0.05}Ca_{0.5}MnO_3$ compounds are referred as PrRE5% (where, RE = La, Y, Eu, Dy, Ho or Tm).

The Rietveld analysis of XRD data reveals that PrRE5% samples form in a single phase orthorhombic structure (space group: *Pnma*, no. 62). The unit cell volume, $\sigma^2$ and $t$ for these samples are listed in Table-I. The parameters, $\sigma^2$ and $t$, are calculated using ionic radii reported for 9 coordination number.[16] It may be noted that $\sigma^2$ increases while $t$ decreases negligibly with decreasing size of RE ions. The decreasing unit cell volume with the decreasing size of $RE^{3+}$ ions further confirms the substitution of $RE^{3+}$ for $Pr^{3+}$ at $A$-site.

Figure 1(a-e) shows magnetization isotherms for $Pr_{0.45}RE_{0.05}Ca_{0.5}MnO_3$ manganites at 1.8 K and 2.5 K. Owing to enhanced $\sigma^2$, these manganites show i) successive sharp MMT(s) from AFM to FM state (with $H_c \sim 8$ T) while sweeping the magnetic-field from zero to 12 T, and ii) a large irreversibility in magnetization while sweeping the magnetic-field from 12 T to zero. On further field cycling, the samples remain in the last acquired magnetic state and do not retrieve the initial magnetic state until warmed up to a temperature above 250 K. This is the indicative of the bi-stable magnetic state at low temperature. However, unlike Ba substituted $Pr_{0.5}Ca_{0.5}MnO_3$ manganites[5] none of the PrRE5% compounds attained the saturated magnetization ($\sim 3.5$ $\mu_B$ for half-doped manganites) up to a magnetic-field of 12 T which may be attributed to the lesser $t$ and $\sigma^2$ of the present system. The unsaturated magnetization state of PrRE5% indicates that the melting of charge-ordered state is partial and that a magnetically phase-separated state remains even at a high field of 12 T. The multiple step-like metamagnetic jumps at 1.8 K smear to smooth transition due to a marginal increase in the temperature to 2.5 K. With further increase in temperature to 5 K, all the sharp MMTs smear to a broad MMT in all these compounds. On comparing the magnetization isotherms of the manganites with nonmagnetic RE ($Eu^{3+}$, $Y^{3+}$) ions with those of manganites with magnetic RE ($Dy^{3+}$, $Ho^{3+}$, $Tm^{3+}$) ions, it is clear that the step-like MMTs can be induced regardless of any intrinsic magnetic property of RE ions. Although these manganites have only marginal difference in their $\sigma^2$, the maximum magnetization attained by PrTm5% at 12 T may be ascribed to its highest cation disorder. In the further discussions, we have defined one magnetic-field cycle as: the magnetic-field sweeping from zero to a maximum magnetic-field ($H_m$) and back from $H_m$ to zero (*i.e.*, zero $\rightarrow H_m \rightarrow$ zero).

Fig. 2 shows resistivity isotherms recorded in magnetic-field cycle of 0 T$\rightarrow$14 T$\rightarrow$0 T at 1.8 K, typically for PrDy5%, PrHo5% and PrY5% samples. The resistivity shows first sharp drop at 8 T and second sharp drop at around 10 T, both of which are consistent with the successive metamagnetic jumps (Fig. 1). Since these samples do not attain a saturation of magnetization (Fig. 1), the sharp drops in resistivity indicate the opening of conducting FM channels in the background of AFM-CO state.[17] On sweeping back the magnetic-field from 14 T to zero, these samples exhibit an irreversible and lower resistivity. This irreversible resistivity is compatible with the irreversible isothermal magnetization (Fig. 1), where the samples retain the induced partial FM phase in this reverse field cycle. This behavior of resistivity is expected for such metamagnetic manganites. However, we observed a contrasting and unexpected behavior of resistivity at high temperature range (above $\sim$70 K) as described in the following text.

The resistance vs. magnetic-field isotherms were measured at various temperatures in the range of 80 – 200 K. The samples were warmed up to room temperature to remove the history effect before starting a new measurement at any temperature. The typical resistivity isotherms are shown for two samples with RE = Eu and Dy at different temperatures (Fig. 3). It is important to note that, in this range of temperature, the magnetization shows reversibility as typically shown for PrDy5% at 80 K and 200 K (insets in Fig. 3). However, the resistivity shows irreversibility with higher values than initial at low magnetic-fields, below 200 K. We quantify the irreversibility of resistivity at zero magnetic-field as follows.

$$\rho_{irr} = (\rho_{cycld} - \rho_{intl}) / \rho_{intl},$$

where, $\rho_{cycld}$ = Resistivity at zero magnetic-field after a magnetic-field cycle of (zero $\rightarrow$ H$_m$ $\rightarrow$ zero), and $\rho_{intl}$ = Initial resistivity at zero magnetic-field before starting a magnetic-field cycle.

The $\rho_{irr}$ for different temperatures is shown in Fig. 4. The $\rho_{irr}$ is found to reduce with the increasing temperature. This irreversible resistivity is caused by the training effect as explained in the following descriptions. As a sample is cooled, it passes through a transition from charge-disordered to a charge-ordered state at ~230 K. At a constant temperature in the range of 80 – 200 K, while sweeping the magnetic-field from zero to 14 T, it passes through a transition from a charge-ordered state to a partially charge-disordered FM state leading to a co-existence of highly competing FM and AFM domains. Now sweeping the magnetic-field back from 14 T to zero, the sample again passes through a transition from partially charge-disordered FM state to AFM-CO state. Thus, on completion of a cycle (zero$\rightarrow$14T$\rightarrow$zero), a part of the sample experiences the repetition of the order-disorder transition. The polycrystalline samples with randomly distributed AFM and FM phases are prone to the training effect on repetition of AFM transition, which may not be observed in single-crystals.[18] The grain boundaries of such polycrystalline samples may contribute to expand the interface regions. As a consequence of the training effect, the AFM phase tries to find energetically favored spin configurations and the AFM-FM exchange bias reduces on repeated magnetic-field cycling.[18-20] A training effect on the resistivity due to thermal cycling across FM transition and on the magnetization due to the magnetic-field cycling, is known where the magnetization is also irreversible.[11-14] However, in the present case, the training effect observed by magnetic-field cycling induced irreversible resistivity is unusual since it manifests in the temperature regime where magnetization is reversible.

Resistivity is a microscopic property which can respond to even local changes occurring in the charge-conduction path. Hence, compared to bulk-magnetization, it is more effective to sense even small changes. As the sample experiences the training effect by a magnetic-field cycle, the resistivity may enhance[21] and show irreversibility even due to slightly favored AFM-CO state. Furthermore, as the temperature increases, the training effect becomes less pronounced and the irreversibility in the resistivity vanishes. This temperature dependence of irreversibility may suggest that the exchange bias reduces as temperature approaches T$_N$.[19] It is established that the training effect disappears around T$_N$.[19] The observation of the disappearance of irreversibility in resistivity in our samples at around T$_N$ (~180 K for undoped Pr$_{0.5}$Ca$_{0.5}$MnO$_3$)[2] confirms that the irreversible resistivity is merely caused by the training effect.

An intriguing question remains: Why was there no change observed in the bulk-magnetization as a result of the training effect as opposed to the others? We give the most plausible reason as follows. In the earlier studies on manganites,[11-14] the training effect was induced i) by thermal cycling in the FM temperature regime or ii) by magnetic-field cycling, both of which resulted in the reduction of 'FM' phase. Furthermore, such reduction of FM content was observed in the manganites with a considerably large fraction of FM phase. Now, a reduction in the 'FM' content can get detected by the bulk-magnetization measurements and for this case, it seems easy to probe the training effect. In the present study, the training effect has been observed at 'high temperatures' and in the 'low' magnetic-field regime amid 'AFM' state. Moreover, our samples attain a low *FM content at high magnetic-fields* before retrieving *AFM state at low magnetic-fields* in a field cycle, or in other words a lesser part of the samples experiences the training effect. This would obviously make it difficult to probe local/small changes in AFM-CO state by bulk-magnetization. However, on magnetic-field cycling, there are modifications occurring at the microscopic level in manganites.[17, 21] As described earlier, the resistivity being microscopic property can sense such changes and hence, the training effect experienced by these samples.

To obtain a better insight on the irreversible resistivity of PrRE5% samples in the high temperature regime, we measured magnetic-field dependent resistivity of two of these manganites in three continuous magnetic-field cycles (*i.e,* 3 cycles of zero→$H_M$→zero) at a fixed temperature of 120 K. In this manner, the resistivity measurements were performed for five different $H_m$ = 3T, 6T, 8T, 10T and 14T. After three continuous cycles for every specific $H_m$, the history effect was removed by warming the samples up to room temperature before starting the new measurements for a different $H_m$. The three continuous typical resistivity isotherms for $H_m$ = 6 T and $H_m$ = 14 T, for PrDy5% and PrLa5% samples are respectively shown in Figs. 5 & 6. The metamagnetic and electronic properties of PrLa5% have been earlier reported elsewhere.[6] The field-cycling effects on both of these manganites are observed to be very similar and further confirm the training effect influenced irreversible resistivity as explained below.

For the constant temperature of 120 K, the resistivity isotherms are discussed in two parts.

*i) The irreversible resistivity in the first field cycle of zero→ $H_m$→ zero; $H_m$ = 3T, 6T, 8T, 10T or 14T:* The resistivity enhances after a magnetic-field cycle up to every specific $H_m$. Fig. 7 shows a correlation of $\rho_{irr}$ and $H_m$ for both PrDy5% and PrLa5% samples. The $\rho_{irr}$ enhances with increasing $H_m$. This may be attributed to the fact that, at a higher magnetic-field ($H_m$), a larger fraction of the sample attains FM state. This FM fraction of the sample retrieves the AFM state while sweeping back the magnetic-field from $H_m$ to zero. Thus, a larger fraction of the sample experiences the training effect at higher $H_m$, giving rise of a larger irreversible resistivity.

*ii) The irreversible resistivity in three continuous field cycles of zero→$H_m$→zero; $H_m$ = 3T, 6T, 8T, 10T or 14T:* As mentioned earlier, the resistivity isotherms are measured in three continuous field cycles for every different $H_m$ (Figs. 5 & 6). For lower $H_m$ (3 T, 6 T and 8 T), there is a small enhancement in $\rho_{irr}$ in the first cycle, but it keeps enhancing on further two cycles. However, in second and third cycles, the enhancement in resistivity is even smaller compared to that in the first cycle. As for higher $H_m$ (10 T and 14 T), there is initially a large enhancement in $\rho_{irr}$ in the first cycle, but in the second and third cycles, a negligibly enhanced $\rho_{irr}$ is observed. However, a common feature is that the training effect becomes less and less pronounced in the successive magnetic-field cycling (for all $H_m$). For the training effect, it is well established fact that the exchange bias reduces with successive magnetic-field cycling.[15,18-20] The decreasing irreversibility of resistivity on successive field cycles complies with this typical character of the training effect.

To summarize, the increase in cation disorder by smaller isovalent RE substitution for Pr in $Pr_{0.5}Ca_{0.5}MnO_3$ induces multiple step-like metamagnetic transitions around 8 T of applied magnetic-field at 2.5 K. At any fixed temperature in the temperature range of 80 K to 170 K, the magnetic-field-cycling influences 'irreversibility' in the resistivity of these manganites in spite of 'reversible' bulk-magnetization isotherms. The irreversible character of resistivity is strongly dependent on the temperature and the magnetic-field. The irreversibility of resistivity gradually decreases as the temperature increases and completely vanishes around $T_N$. This along with a decrease in irreversibility of resistivity on successive field cycling, are understood to be originated due to a training effect arising as a result of magnetic-field-cycles. These results further suggest that a microscopic study may give clear insight into the AFM-FM exchange interactions and training effect on such manganites.

Table I: Tolerance factor ($t$), cation size disorder ($\sigma^2$) and unit cell volume for $Pr_{0.45}RE_{0.05}Ca_{0.5}MnO_3$ (PrRE5%) samples.

| Sample | $t$ | $\sigma^2$ ($10^{-4} Å^2$) | Unit cell volume ($Å^3$) |
|---|---|---|---|
| PrEu5% | 0.9166 | 1.7 | 220.71(3) |
| PrDy5% | 0.9160 | 4.4 | 220.51(3) |
| PrY5% | 0.9159 | 5.2 | 220.50(3) |
| PrHo5% | 0.9158 | 5.5 | 220.48(3) |
| PrTm5% | 0.9155 | 7.7 | 220.27(3) |

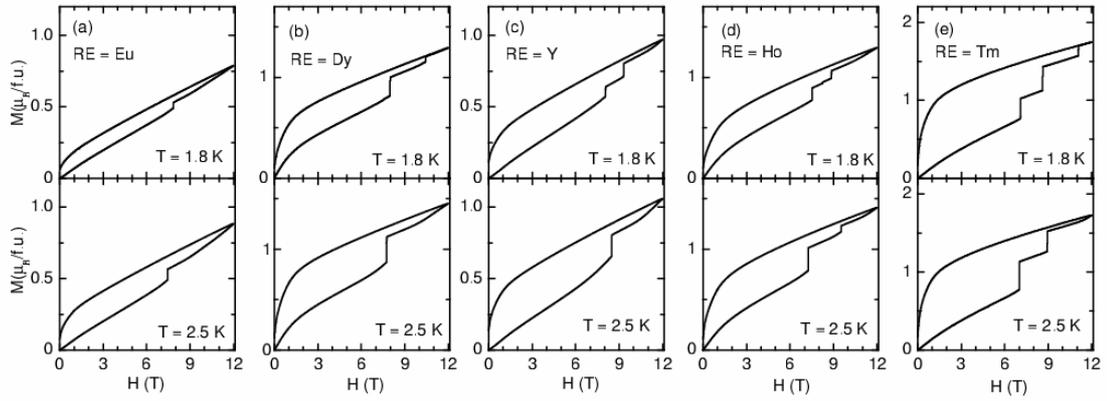

Figure1(a-e): Magnetization isotherms for $Pr_{0.45}RE_{0.05}Ca_{0.5}MnO_3$ manganites at 1.8 K and 2.5 K.

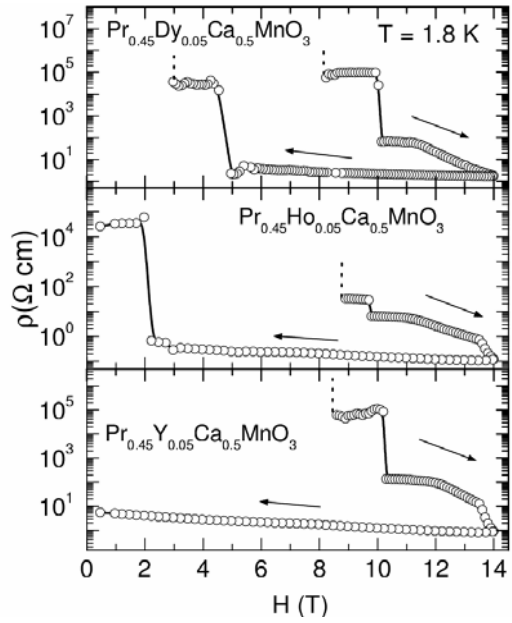

Figure 2: Typical resistivity isotherms for three $Pr_{0.45}RE_{0.05}Ca_{0.5}MnO_3$ (RE = Dy, Ho, Y) manganites at 1.8 K.

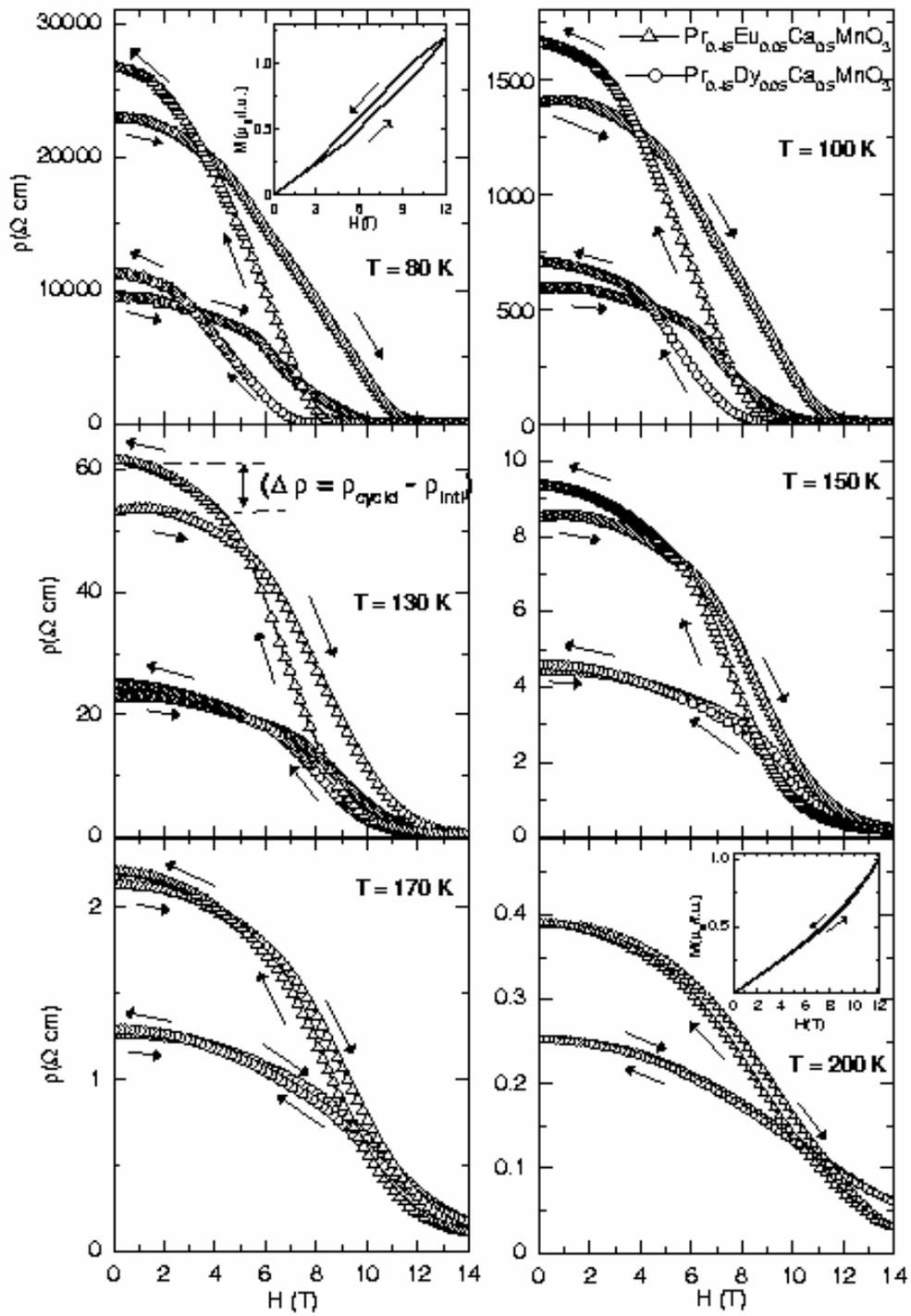

Figure 3: Resistivity isotherms for $Pr_{0.45}RE_{0.05}Ca_{0.5}MnO_3$ (RE = Eu, Dy) manganites in a temperature regime of 80 K to 200 K. The inset figures show typical magnetization isotherms for PrDy5% at 80 K and 200 K. The irreversibility in resistivity is defined as, $\rho_{irr} = \Delta\rho / \rho_{intl,}$ where $\Delta\rho = \rho_{cycld} - \rho_{intl,}$ as shown in the figure.

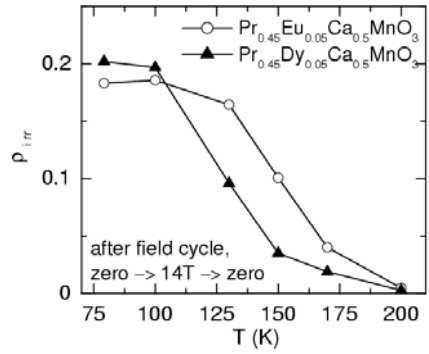

Figure 4: Dependence of irreversibility of resistivity on the temperature from 80 K to 200 K.

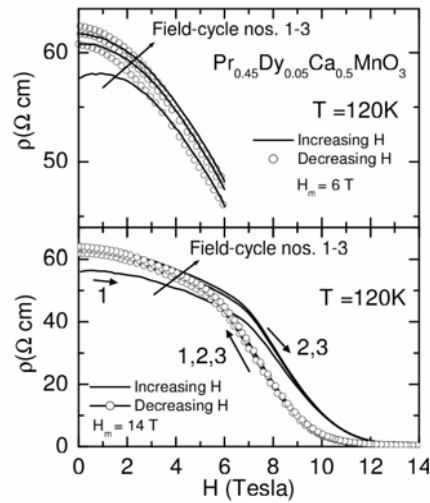

Figure 5: Typical resistivity isotherms in repeated magnetic-field-cycles up to two different magnetic-fields ($H_m$ = 6 T and 14 T), for $Pr_{0.45}Dy_{0.05}Ca_{0.5}MnO_3$ at 120 K.

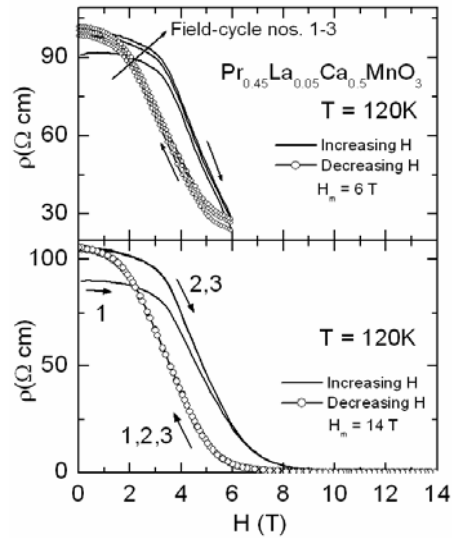

Figure 6: Typical resistivity isotherms in repeated magnetic-field-cycles up to two different magnetic-fields ($H_m$ = 6 T and 14 T), for $Pr_{0.45}La_{0.05}Ca_{0.5}MnO_3$ at 120 K.

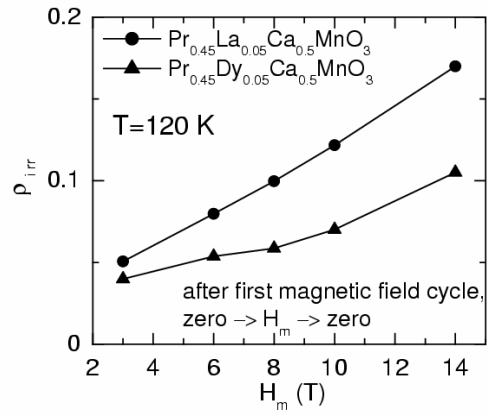

Figure 7: Dependence of irreversible resistivity on $H_m$ after one magnetic-field cycle (zero $\rightarrow$ $H_m$ $\rightarrow$ zero) at 120 K.

*****